\documentclass{ws-ijmpb}

\usepackage{graphicx,psfrag,feynmf}

\begin{document}

\markboth{A. Virosztek, K. Maki and B. D\'ora}
{Micromagnetism in URu$_2$Si$_2$ and HTSC}

\catchline{}{}{}

\title{MICROMAGNETISM IN URu$_2$Si$_2$ AND\\
HIGH TEMPERATURE SUPERCONDUCTORS}

\author{\footnotesize ATTILA VIROSZTEK}

\address{Department of Physics, Technical University of Budapest, H-1521
Budapest, Hungary, and\\
Research Institute for Solid State Physics and Optics, P.O.Box 49, H-1525
Budapest, Hungary}

\author{KAZUMI MAKI}

\address{Department of Physics and Astronomy, University of Southern
California, Los Angeles, CA 90089-0484, USA}

\author{BAL\'AZS D\'ORA}

\address{Department of Physics, Technical University of Budapest, H-1521
Budapest, Hungary}

\maketitle

\pub{Received (received date)}{Revised (revised date)}

\begin{abstract}
It has been proposed, that unconventional density waves (UDW) are possible
candidates for systems with hidden order parameter. Unlike in conventional
density waves, no periodic modulation of either the charge-, or the
spin-density is present in UDW, in spite of a clear thermodynamic signal.
Although the unconventional spin density wave (USDW) has been suggested for
the "antiferromagnetic" phase of URu$_2$Si$_2$, the micromagnetism seen by
neutron scattering has not been understood. We present here the calculation
of the local spin density due to impurities in USDW, which describes
quantitatively the neutron scattering data by Amitsuka {\it et.al.}. Further,
we propose that the pseudogap phase in high temperature superconductors
(HTSC) should also be USDW. Strong evidence for this are the micromagnetism
seen by Sidis {\it et.al.}, and the optical dichroism seen by Campuzano
{\it et.al.}.
\end{abstract}

\section{Introduction}

Many unusual
features of the high temperature superconductors (HTSC)
are understood in terms of a momentum dependent
order parameter $\Delta({\bf k})$. It follows
from the success of these developments naturally, that investigating the
properties of
unconventional condensates in the density wave sector may as well turn out to
be a fruitful enterprise. Indeed, in the last few years researchers
encountered
mysterious low temperature phases in a number of materials, where a clear
and robust thermodynamic phase transition is not accompanied by an order
parameter detectable by conventional means. This situation is often referred
to as hidden order.

Prime example is the 17.5~K transition in URu$_2$Si$_2$
with very small magnetic moment in the low temperature
phase. It has been suggested\cite{IO}, that an unconventional spin density
wave (USDW) may be responsible for this behavior, since in the clean system
USDW does not exhibit periodic spin density modulation due to the vanishing
momentum average of the order parameter. The origin of the small but finite
magnetic moment, termed micromagnetism, however remains elusive. The magnetic
phase diagram of some $\alpha$-(ET)$_2$ salts also offers a low temperature
phase (LTP) with resemblance to charge density waves (CDW)\cite{Christ}, but
without the characteristic X-ray satellites. In this situation a quasi-one
dimensional UCDW scenario\cite{DV} is called for, and it's case is
strengthened by the evaluation of the corresponding threshold electric field
for sliding conductivity\cite{PRBRC}. Finally, the pseudogap phase of high
temperature superconductors was proposed to be a kind of UCDW as
well\cite{Benfatto,Sudip}. However, micromagnetism was also observed in YBCO
by neutron scattering\cite{Sidis}.

In this paper we propose a mechanism by which the USDW exhibits weak priodic
spin density modulation in the presence of random magnetic impurities.
Comparing the temperature dependence of neutron scattering intensity due to
this modulation with experimental data on URu$_2$Si$_2$\cite{Ami}, we argue
that this mechanism is responsible for the micromagnetism in this material,
and possibly also in the pseudogap phase of HTSC.

\section{General features of unconventional density waves}

The unconventional spin density wave (USDW) with spin polarization in the
$\alpha$ direction ($\alpha=x,y,z$) is described in the Nambu
formalism by the Green's function
$
G^{-1}({\bf k},i\omega_n)=i\omega_n-\rho_3\xi({\bf k})+\rho_1\sigma_\alpha
\Delta({\bf k})
$,
where the $\rho$ and $\sigma$ Pauli matrices operate on the electron-hole,
and spin space respectively, $\xi({\bf k})$ is the electron spectrum measured
from the Fermi energy, and $\Delta({\bf k})$ is the momentum dependent order
parameter. In the quasi-one dimensional case\cite{DV} it can be for example
$\Delta({\bf k})=\Delta\cos(bk_y)$, while for URu$_2$Si$_2$
$\Delta({\bf k})=\Delta[\cos(ak_x)-\cos(ak_y)]$ has been proposed\cite{IO}.
Clearly, due to the vanishing momentum average of $\Delta({\bf k})$, the
expectation value of any component of the spin density is zero\cite{IO,DV}.
It turns out, that the quantity which does have a periodic modulation with
the nesting vector ${\bf Q}$, is the spin current density\cite{DV}.
Nevertheless, in the ideal system there is no spin density modulation,
therefore the micromagnetism remains unexplained, although the Green's
function
definitely prefers a unique direction ($\alpha$)
in spin space. There is no
such spin direction for UCDW, therefore the mechanism for micromagnetism
proposed below will not work in that case.

\section{Local spin density and micromagnetism}

\newcommand{\BM}[1]{\mbox{\boldmath$#1$}}

\begin{figure}[h!]
\begin{fmffile}{uru}
\hspace*{-3cm}
\parbox{80mm}{\begin{fmfgraph*}(80,80)
 \fmfleft{a}
 \fmfright{b,c}
 \fmf{phantom}{a,v}
 \fmf{phantom}{v,b}
 \fmf{phantom}{v,c}
 \fmfv{decor.shape=cross,decor.size=4thick}{v}
 \fmf{fermion,left=2,label=${\bf k},,i\omega_n$}{v,v}
 \fmflabel{\hspace*{-2.2cm}$({\bf S}_j\cdot\BM{\sigma})J_{\bf k,k}$}{v}
 \end{fmfgraph*}}\hspace*{2cm}
\parbox{30mm}{\begin{fmfgraph*}(30,22)
 \fmfleft{a}
 \fmfright{b}
 \fmf{fermion,left=0.6,label=${\bf k},,i\omega_n$}{a,b}
 \fmf{fermion,left=0.6,label=${\bf k^\prime},,i\omega_n$}{b,a}
  \fmfv{decor.shape=cross,decor.size=4thick}{b}
 \fmfdot{a}
 \fmflabel{$\BM{\sigma}$}{a}
 \fmflabel{$({\bf S}_j\cdot\BM{\sigma})J_{\bf k,k^\prime}$}{b}
  \end{fmfgraph*}}
 \end{fmffile}
\vspace*{-2.5cm}
\caption{Diagrams contributing to the thermodynamic potential and to the
spin density respectively, in first order in the impurity potential.
\label{fig:graph}}
\end{figure}
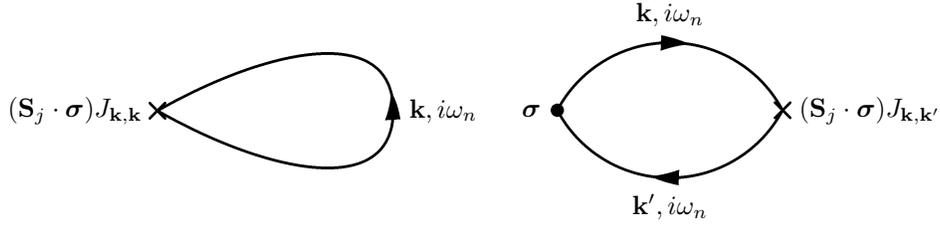

We assume that we have magnetic impurities in our sample at positions
${\bf R}_j$ with spin ${\bf S}_j$ interacting with the electron spin density
through a Heisenberg exchange coupling $U({\bf r})=J({\bf r}-{\bf R}_j)
{\bf S}_j\BM{\sigma}$. The first diagram on Fig.~\ref{fig:graph} shows the
change in the thermodynamic potential due to impurities:
\begin{equation}
\Delta\Omega=T\sum_{\omega_n}\sum_{\bf k}{\rm Tr}[U({\bf k},{\bf k})G({\bf k},
i\omega)],\label{TDpot}
\end{equation}
where $U_{s,s^\prime}({\bf k},{\bf k}^\prime)=\int d^3 r \varphi^*_{{\bf k}+
s{\bf Q}/2}({\bf r})U({\bf r})\varphi_{{\bf k}^\prime+s^\prime{\bf Q}/2}
({\bf r})$ is the matrix element of the impurity potential between Bloch
states ($s,s^\prime=\pm 1$ spanning the electron hole space).
$\Delta\Omega$ vanishes for any kind of CDW, but for USDW each
impurity feels a potential energy $\Delta\Omega=S_j^\alpha(J\Delta/P)\cos
({\bf QR}_j)$, trying to align the inpurity spin parallel to the spin
direction favoured by the USDW, including its periodicity given by ${\bf Q}$.
$P$ is the component of the interaction driving the USDW transition, while
$J$ is the coefficient of the term originating from $U({\bf k},{\bf k})$
(a matrix element of the exchange integral $J({\bf r})$ between Bloch
states),
having the same momentum dependence as the order paprameter does (for example
$\cos(ak_x)-\cos(ak_y)$). 

\unitlength=1mm
\vspace*{-2.5cm}

\begin{figure}[h!]
\psfrag{x}[t][b][1][0]{$T/T_c$}
\psfrag{y}[b][t][1][0]{$I/I_0$}
\vspace*{-0.5cm}
{\includegraphics[width=12cm,height=6cm]{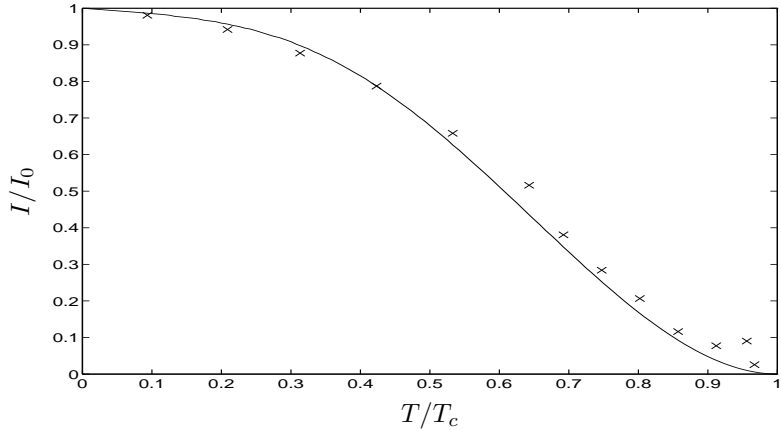}}
\caption{Normalized neutron scattering intensity as a function of reduced
temperature. Crosses denote URu$_2$Si$_2$ data (Amitsuka {\it et.al.}),
the solid line is the theoretical prediction from Eq.(\ref{Ampl}).
\label{fig:urusi}
}
\end{figure}

Once the impurity spins are coherently aligned much like in a conventional
SDW, the electron spin polarization around them will also develop coherently,
facilitating the observation of a magnetic order by neutron scattering. The
electron spin density induced by impurities can be calculated from the second
diagram on Fig.~\ref{fig:graph} as
\begin{equation}
\langle\BM{\sigma}({\bf r})\rangle=T\sum_{\omega_n}\sum_{{\bf k},
{\bf k}^\prime}{\rm Tr}[\BM{\sigma}_{{\bf k}^\prime,{\bf k}}({\bf r})
G({\bf k},i\omega)U({\bf k},{\bf k}^\prime)G({\bf k}^\prime,i\omega)],
\label{Spin}
\end{equation}
where $[\BM{\sigma}_{{\bf k}^\prime,{\bf k}}({\bf r})]_{s^\prime,s}=
\varphi^*_{{\bf k}^\prime+s^\prime{\bf Q}/2}({\bf r})\BM{\sigma}
\varphi_{{\bf k}+s{\bf Q}/2}({\bf r})$. A similar expression has already
been evaluated\cite{PRBRC} for the impurity pinning potential in UDW, which
allows us to express the electron spin density at the impurity site
${\bf R}_j$. The excess spin polarization due to the USDW turns out to be
parallel to the USDW's preferred spin direction ($\alpha$) as well, and the
temperature dependence of its magnitude normalized to the $T=0$ value is
given by
\begin{equation}
{\sigma(T)\over\sigma(0)}=\frac{\Delta(T)}{\Delta(0)}\frac{1}{0.5925}
\int_0^1\tanh\frac{\beta\Delta(T)x}{2}E(\sqrt{1-x^2})(K(x)-E(x))dx.
\label{Ampl}
\end{equation}
The neutron scattering intensity is proportional to the square of this
amplitude, and its normalized temperature dependence is plotted on
Fig.~\ref{fig:urusi}, along with the experimental data\cite{Ami} on
URu$_2$Si$_2$.

\section{Concluding remarks}

We have proposed a mechanism for the development of micromagnetism in
USDW, and shown that the temperature dependence of the magnetic moment is
consistent with measurements on URu$_2$Si$_2$, lending further support for the
suggestion that the low temperature phase of this material is a USDW. We also
believe that the same mechanism may apply for the pseudogap phase in
HTSC\cite{Sidis}, suggesting that it is not UCDW, but USDW instead. This is
further corroborated by the optical dichroism observed in this
phase\cite{Campu}.

\section*{Acknowledgements}

We thank B. Keimer and H. Amitsuka for stimulating discussions.
This work was supported by the Hungarian Scientific Research Fund
under grant numbers OTKA T032162 and T029877, and by the
Ministry of Education under grant number FKFP 0029/1999.

\end{document}